\documentstyle[11pt,paspconf]{article}

\begin{document}

\title{The nature of the UV-optical continuum in Seyfert 2 galaxies}

\author{Thaisa Storchi-Bergmann\altaffilmark{1} and Henrique R.
Schmitt\altaffilmark{2}}
\affil{Instituto de F\'\i sica, UFRGS, Campus do Vale, C.P. 15051, CEP\,91501-970, Porto Alegre, RS, Brasil}

\author{Roberto Cid Fernandes}
\affil{Departamento de F\'\i sica, CFM-UFSC, Campus Universit\'ario Trindade,
       CP\,476, CEP\,88040-900,
       Florian\'opolis, SC, Brasil}


\altaffiltext{1}{Visiting Astronomer, Cerro Tololo Inter-American Observatory. 
CTIO is operated by AURA, Inc.\ under cooperative agreement with the National
Science Foundation} 
\altaffiltext{2}{Presently at the Space Telescope Science Institute}



\begin{abstract}

The nature of the UV-Optical continuum in Seyfert 2 galaxies is a
matter of current debate, fundamental to which is the issue of
characterizing and quantifying the stellar population contribution to
the nuclear spectra. Using high S/N long-slit spectroscopy of a sample
of 20 Seyfert 2 galaxies, we have applied a novel approach to investigate
the nuclear stellar population in these galaxies.
Our main results are: (1) the stellar populations in 
Seyfert 2 galaxies are varied, and in most cases {\bf cannot}
be adequately represented by an elliptical galaxy template,
as done in previous works; (2) the central kpc
of most Seyfert 2's contain substantially larger proportions of 100\,Myr
stars than either elliptical galaxies or normal spirals of the same
Hubble type.  One important consequence of our findings is
that the controversial nature of the so called ``second''
featureless continuum (FC2) in Seyfert 2s  is most likely a
result of inadequate evaluation of the stellar population.

\end{abstract}




\section{Introduction}

The nature of the UV-Optical continuum of Seyfert 2 galaxies has long
been a matter of investigation. If, in one hand, these galaxies are
known to present a bluer continuum than ``normal'' galaxies of the same
Hubble type (this characteristic allowed them to be found in the
Byurakan survey, for example), on the other hand, in the Unified Model
(Antonucci \& Miller 1985, Antonucci 1993), a Seyfert 1 nucleus,
comprising the the nuclear source and broad-line region, is hidden by an
obscuring dusty molecular torus. In this scenario, the nuclear 
featureless continuum (FC) and broad lines are only seen via scattered light,
as indeed observed in a number of polarimetric
studies (Miller \& Goodrich 1990; Kay 1992, Tran 1995).

Nevertheless, Tran (1995) observed that,
after subtraction of the stellar population contribution 
(usually about 70-80\% at $\lambda$5500\AA ) the polarization in the
continuum was smaller than in the broad lines. To reconcile
the two polarization values,
he concluded that there were two ``featureless'' continua
components: FC1, consisting of scattered light from the nuclear source,
contributing with typically 5\% to the total observed continuum, and
FC2, an unpolarized component contributing with the remaining 15-25\%
of the continuum.

This study raised the new issue of the nature of the FC2 component. One
possibility first proposed by Cid-Fernandes \& Terlevich (1992, 1995)
was the contribution from young stars in a nuclear burst of star
formation. A handful of Seyfert 2's is  known to have composite
nuclei, classified as Starburst + Seyfert (Kinney et al. 1993), and
recent studies (Heckman et al. 1997; Gonzalez-Delgado et al. 1998) have
shown that the signatures of the starburst in these cases are evident
in the UV-blue nuclear spectra of the galaxies. In these few Seyfert
2's, FC2 could indeed be identified with young ($< 10$\ Myr) stars.  
Another possibility, proposed by Tran (1995), based on previous results
by Barvainis (1993), was that FC2 is due to thermal free-free emission
from a hot plasma wind.  Recently, Axon, Capetti \& Macchetto (1998) argued
that the blue continuum can be due to free-free
emission from gas shocked by the passage of a radio jet.

In this paper we discuss the results of our recent work, 
dedicated to investigate the nuclear stellar population and possible 
additional continuum in Seyfert 2 galaxies. We have used a novel
approach which combines a spatially resolved study of the
stellar population and spectral synthesis of the nuclear spectra.

\section{The Stellar Population in Active Galaxies}

In order to investigate the stellar content of Seyfert galaxies,
as well as to compare it with that of other galaxies,
Cid-Fernandes, Storchi-Bergmann \& Schmitt (1998), have analized   
long-slit optical spectra (obtained with the CTIO 4m telescope) of
20 Seyfert 2's, 6 Seyfert 1's, 7 LINER's,
5 radio-galaxies, 1 elliptical and 3 "normal" galaxies with
nuclear rings of star formation.

Our approach was to measure the continuum flux in selected windows and
the equivalent widths (W's) of several stellar absorption features
as a function of distance from the nuclei, at a spatial sampling of
$2^{\prime\prime}\times 2^{\prime\prime}$. 

\begin{figure}
\vspace{1.75in}
\includegraphics{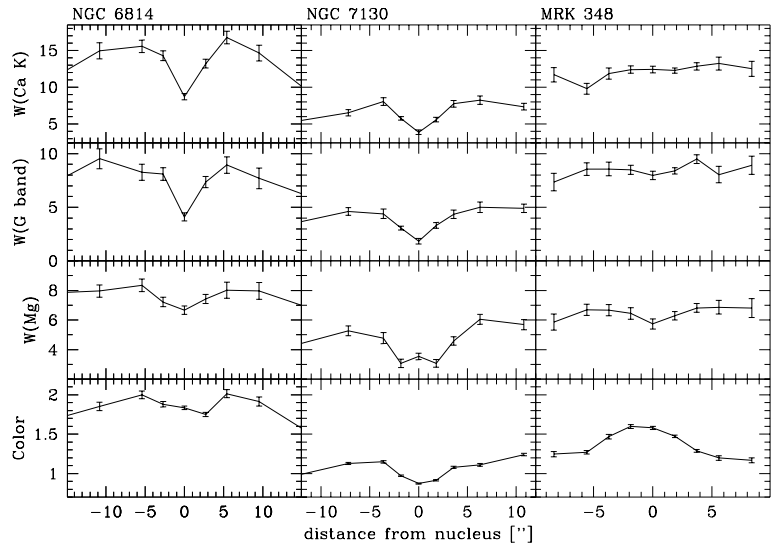}
\caption{From top to bottom: Radial variations of the equivalent 
widths of the absorption features Ca II K ($\lambda$3930\AA), 
G-band ($\lambda$4301\AA), Mg I + MgH ($\lambda$5176\AA) and
continuum ratio $\lambda$5870/$\lambda$4020 for the Seyfert 1
galaxy NGC\,6814 (left), the composite Seyfert $+$ Starburst
galaxy NGC\,7130 (center) and the Seyfert 2 Mrk\,348 (right). (Adapted
from Paper I).} \label{fig-1}
\end{figure}

The main conclusions of this work were:  
(1) there is a large diversity of nuclear
stellar populations characteristics among Seyfert galaxies, and in most
cases, differ from those of an elliptical galaxy; 
(2) the W's in regions of star-formation
(e.g., star-forming rings), in the nuclear spectra of Seyfert 1's
and for the Seyfert 2 galaxies classified as composite (SB+Sy 2) are
observed to be smaller than in nearby regions, indicating a dilution by
an underlying continuum, as expected; (3) for most Seyfert 2's (17 out
of 20), we found no dilution in the nuclear EW's, leading to the
conclusion that if there is a FC continuum present, it contributes {\it
at most} with 10\% at all optical wavelengths.  Conclusions (2) and (3)
are illustrated in Figure~\ref{fig-1}.

%

\section{The Nature of Optical Light in Seyfert 2 Galaxies 
with Polarized Continuum}

In a subsequent work, Storchi-Bergmann, Cid-Fernandes \& Schmitt (1998)
look closely at the spectra of 4 Seyfert 2's with no dilutions 
in the nuclear stellar absorption W's, but with previous determinations of FC contributions larger than 10\%. The 4 galaxies are:
{\it Mrk 348}, with 27\% contribution at 5500\AA\ of a nuclear FC (Tran 1995): 
from polarimetry, it was concluded that 5\% was due to the polarized
continuum (FC1), but  22\% was due to FC2; {\it Mrk 573}, 
with  20\% contribution at 4400\AA\ of a nuclear FC (Kay 1994);
{\it NGC 1358}, also with 20\% contribution of FC at 4400\AA\ (Kay 1994);
and {\it Mrk 1210}, with 25\% contribution at 5500\AA\ (Tran 1995): 
from polarimetry, 6\% of FC1 and 19\% of FC2.

In both Kay's (1994) and Tran's (1995) work, it was assumed that the 
nuclear stellar population was well represented by the spectrum
of a typical elliptical galaxy.

The above contributions of a nuclear FC seem to be in contradiction
with the results of Cid Fernandes et al. (1998), if: 
(1) FC is confined to the nucleus (inner 2$^{\prime\prime}\times
2^{\prime\prime}$); (2) the stellar population does not vary within
the bulge of the galaxy -- it was verified that the bulges have
average effective radius of 6$^{\prime\prime}$ for the sample. 
If these two conditions are true, then the results of 
Cid Fernandes et al. (1998) indicate
that the FC contribution to the nuclear spectra should be smaller
than 10\%, in contradiction with the results of previous works.

Storchi-Bergmann et al. (1998) then adopted the first hypothesis
above  and tested the second,
extracting spectra from the bulge of the same galaxy from windows at 4$^{\prime\prime}$ from the nucleus,
which were used as stellar population templates for the nuclear spectra.

\begin{figure}
\vspace{1.75in}
\includegraphics{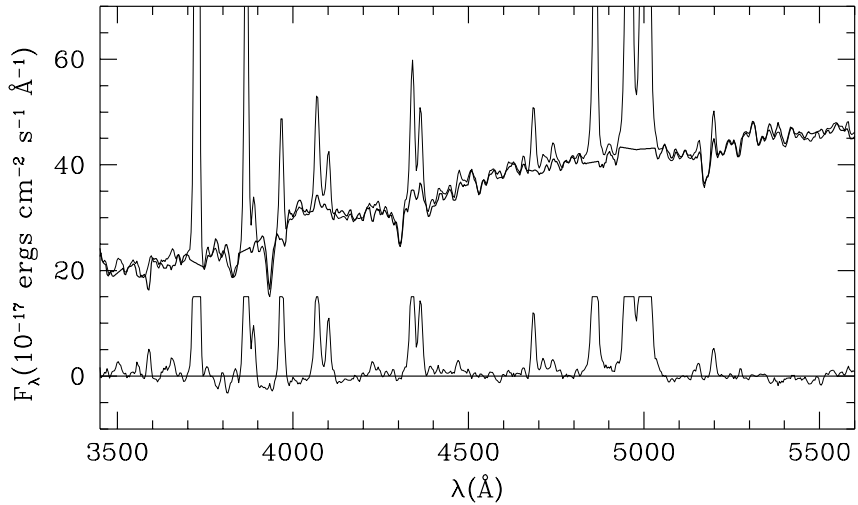}
\caption{The Mrk\,348 nuclear spectrum (thin line) is compared with the
extranuclear spectrum reddened by E(B-V)=0.09, combined with
5\% contribution (at $\lambda$5500\AA) of the FC1 component,
represented by a $F_\lambda \propto \lambda^{-1}$ power-law (thick line). 
The residual between the two is plotted at the bottom, where the emission lines
have been chopped for clarity.} \label{fig-2}
\end{figure}

These extranuclear spectra were then used in combination with a
small percentage ($\approx$5\%) of FC1 (represented by a power-law
spectrum) as derived by Tran (1995) to reproduce the nuclear spectrum.
It was concluded that, after applying some reddening to the extranuclear
spectra,  this combination provided a good representation
of the nuclear spectrum, with no need of FC2, as illustrated
in Figure~\ref{fig-2} for Mrk\,348.

Population synthesis of these extranuclear spectra then showed that 
an elliptical galaxy template was only valid for Mrk\,573, with
the other three galaxies presenting larger proportions of
younger components. It was thus concluded that the need of
FC2 in these cases was a consequence of the fact that an elliptical
template is not adequate  to represent the stellar population of 
most Seyferts.

\section{Spectral Synthesis of the Nuclear Region of Seyfert 2 Galaxies}

In order to investigate the above result for the whole sample of Seyfert 2
galaxies, Schmitt, Storchi-Bergmann \& Cid Fernandes (1998) 
performed spectral syntheses for
the nuclear spectra of the 20 Seyfert 2's, comparing the results
with those of an elliptical galaxy template. The spectral base
comprised  star cluster spectra 
of different ages (Bica, 1988), an HII continuum and a power-law
component to represent the scattered component (FC1).

The results can be summarized through the 
contribution of the different age components
to the light at $\lambda$5780\AA\ for the nuclear Seyfert 2 spectra
compared with that for the elliptical template:
19/24 (80\%) have larger contributions of stars of 100 Myrs;
7/24  (30\%) have larger contributions of stars of 10 Myrs;
5/24 (20\%) have larger contributions of an HII continuum.

It is also interesting to compare the above results with those
for the bulge of early type spirals, the dominant Hubble types
of the Seyfert 2's. From Bica (1988), it can be concluded that in
only 20\% of a sample of 51 early type spirals, there are contributions
of stars of 100 Myrs or younger.
It can be concluded that {\it the main difference between the stellar population
of Seyfert 2's and of elliptical or early-type spiral galaxies
is the large contribution of stars with 100 Myrs}.

\section{Conclusions}

The main conclusions of this work can be summarized as follows.

(1) Galaxies with Seyfert nuclei have a varied nuclear stellar population.

(2) In most cases this population differs from that of  an   elliptical 
galaxy. When a proper template is used, preferably from the bulge of
the same galaxy, there is no need of  FC2. Alternatively,
if FC2 is identified with the difference between the unpolarized continuum
and the elliptical template, it is due to stars with ages 0-100 Myr.

(3) The main  difference between the nuclear stellar population of 
Seyferts and of early-type spirals and ellipticals is the
larger contribution of a 100 Myr population.

(4) We could speculate that if there is a causal 
link between star-formation and nuclear activity, the 100 Myr star-formation timescale should be comparable to duration of nuclear activity cycle. Since 100 Myr is $\approx$ 1\% of the Hubble time, $\approx$1\% of the galaxies 
should be active, in agreement with recent estimates (e.g., 
Huchra \& Burg 1992).

%
%

%

\end{document}